\documentclass[12pt]{iopart}

\usepackage{iopams} 
\usepackage{mathrsfs}

\usepackage{graphicx}
\usepackage{amstext}
\usepackage{bm}
\usepackage{amsfonts}
\usepackage{epstopdf}
\usepackage{cite}

\expandafter\let\csname equation*\endcsname=\relax
\expandafter\let\csname endequation*\endcsname=\relax
\usepackage{amsmath,amssymb,amsthm}

\begin{document}

\title[]{Two-tone modulated cavity electromagnonics}

\author{Nianqi Hu and Huatang Tan* }

\address{ Department of Physics, Huazhong Normal University, Wuhan 430079, China}
\ead{tht@mail.ccnu.edu.cn}
\vspace{10pt}

\begin{abstract}
Cavity electromagnonics has increasingly emerged as a new platform for the fundamental study of quantum mechanics and quantum technologies. Since the coupling between the microwave field and magnon Kittle modes in current experiments is much weaker than their resonant frequencies, the anti-rotating terms in magnon-microwave-photon interaction can be neglected and only the beam-splitter-like part takes effect. In this situation, the direct generation of magnonic nonclassical states is impossible, unless other subsystems e.g. phonons, squeezed photons or superconducting qubits are incorporated. In this paper, we consider two-tone modulated cavity electromagnonics to keep the nontrivial anti-rotating terms and obtain tunable phase factors, resulting in an effective Hamiltonian exactly the same as that of generic linearized cavity optomechanics. This can therefore be exploited to directly prepare macroscopic magnonic quantum states, as detailedly exemplified by the generation of steady and strongly squeezed and entangled states, realize ultra-sensitive magnon-based sensing by engineering backaction-evading interaction of magnons and photons, and develop spintronics-related quantum information processing devices.
\end{abstract}

%
\vspace{2pc}

\section{Introduction}
Recently, exploring quantum effects in hybrid systems based on magnons (collective spin excitations in magnetic materials) has attracted increasing attention \cite{ia1, ia2, ia3}. This is because magnonic systems, like yttrium iron garnet (YIG) spheres with diameter around hundreds of micrometers, can be a good candidate for studying macroscopic quantum phenomena \cite{mqp1, mqp2, mqp3, mqp4} and are of importance for magnon-based quantum technologies, such as quantum information processing\cite{qip}, quantum sensing\cite{qs1, qs2}, and quantum networks \cite{qn}. Hybrid magnonic systems also possess excellent qualities, such as great tunability of frequency \cite{tof} as well as low dissipation rates \cite{ldr1, ldr2, ldr3, ldr4},
which provides efficient way to control magnonic states.
Experimentally, resolving magnon Fock states by coupling of
magnons in a YIG sphere to a superconducting quantum qubit
has been achieved \cite{sqq1,sqq2,sqq3}. Recent studies
have also revealed that nonclassical properties, such as quadrature
squeezing \cite{aqs1, aqs2, aqs3, aqs4, aqs5}, entanglement \cite{ae1, ae2, ae3, ae4, ae5, ae6, ae7}, and magnon quantum blockade \cite{mqb1, mqb2, mqb3, mqb4}, can be generated in magnonic systems.

Cavity electromagnonics focus on the interaction between microwave photons and magnons. Since nearly resonant frequencies of microwave photons and magnons, strong electromagnonic coupling, compared to the loss rates of microwave cavity and magnonic modes, has been experimentally realized \cite{ldr1, ldr3, se0, se1, se2, se3, se4, se5}.
This enables novel phenomena and applications, including non-Hermitian physics \cite{inp1, inp2, inp3}, microwave-to-optical transduction \cite{mot1, mot2}, and coherent gate operations \cite{cgo1}. Nevertheless, the magnon-microwave photon coupling in current experiments is on the order of hundreds of megahertz, far away from the resonances of the magnon and microwave photons at tens of gigahertz. As a result, the counter-rotating part, taking the form of magnon-photon parametric downconversion,  in the magnetic-diploe interaction between magnons and microwave photons take little effect and can thus be neglected, and only the near resonant part, i.e., beam-splitter-like interaction is kept. The latter part is just responsible for the coherent exchange between the photon and magnons and therefore in this situation the nonclassical magnonic or microwave states are impossibly achieved directly by cavity electromagnonics. To achieve the nonclassical states such as squeezing and entanglement, other subsystems, e.g., phonons \cite{ps1, ps2}, squeezed photons \cite{ae3, sps1},  superconducting qubits \cite{sq1, sq2, sq3} or weak Kerr nonlinearity \cite{wkn1, wkn2, wkn3, wkn4} should be incorporated to the cavity electromagnonic systems, which unavoidably brings out excess noise and just leads to weak squeezing and entanglement. A question naturally arise: Can we keep the counter-rotation terms and obtain simultaneous magnon-microwave photon parametric downconversion and beam-splitter-like interactions to generate various magnonic quantum states and implement quantum tasks, in analogy to cavity optomechanics in which fruitful results have been achieved in the past decades \cite{co}?

In this paper, we consider a two-tone modulated cavity electromagnonics in which the magnonic frequency is modulated by external two-tone magnetic field to keep the nontrivial anti-rotating terms. We note that one-tone Floquet cavity electromagnonics has been domenstrated \cite{fce1} and it can be used for chiral magnon current \cite{cc1}. It is shown that through the two-tone modulated cavity electromagnonics, an effective Hamiltonian exactly same as that of generic linearized cavity optomechanics can be obtained, which allows us to engineer nonclassical magnonic states directly via cavity electromagnonics, in analogy with cavity optomechanics. The modulation can also be used to realize back-action-evading measurement of magnonic amplitude for magnon-based quantum metrology. The scheme provides promising opportunities for the preparation of macroscopic quantum states of collective spin excitations in solids and magnon-based quantum information processing and metrology.


\section{Two-tone modulated electromagnonical system}

\begin{figure}[t]
\centerline{\scalebox{0.35}{\includegraphics{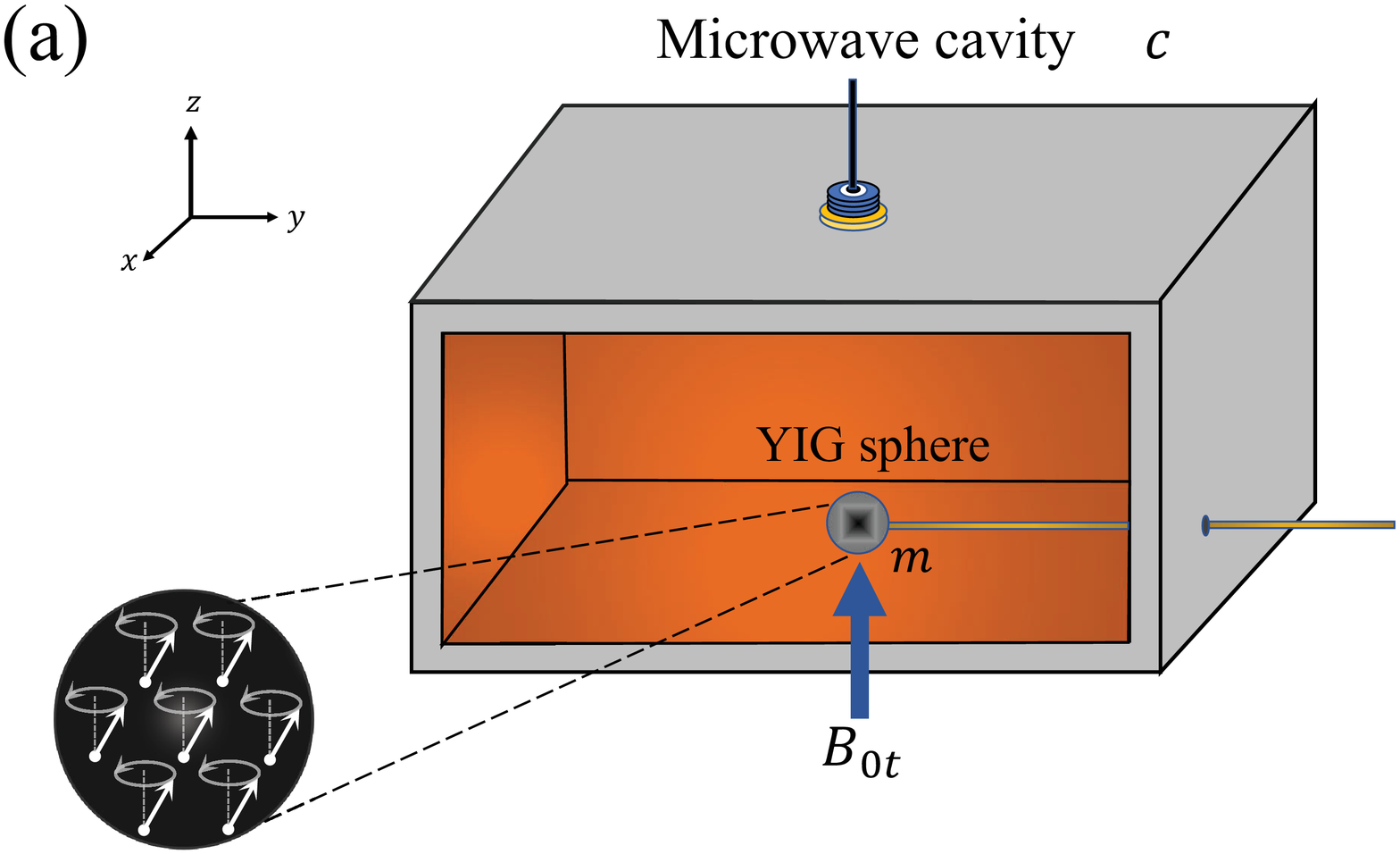}}}
\vspace{0.01cm}
\centerline{\scalebox{0.35}{\includegraphics{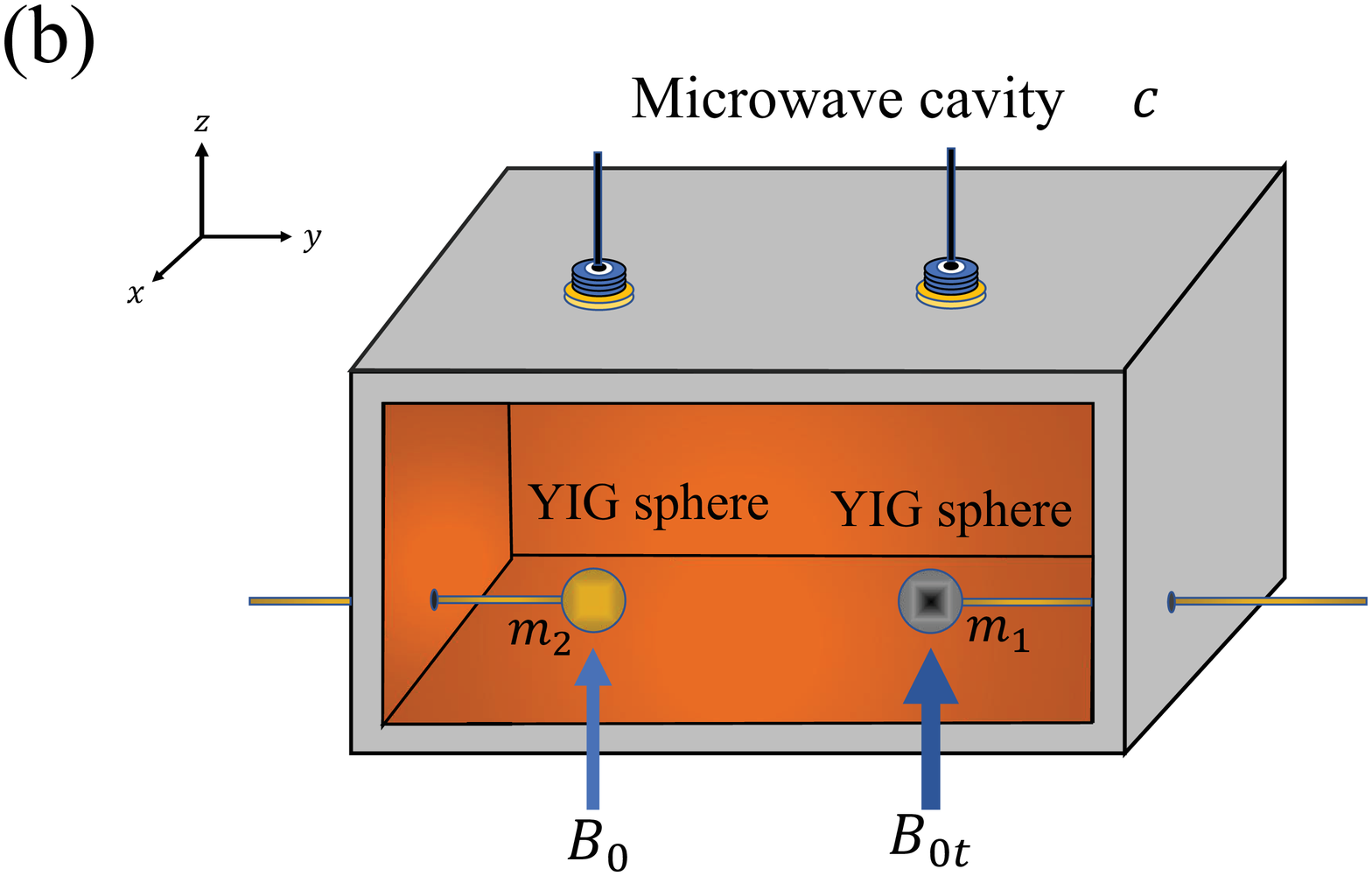}}}
\vspace{0.01cm}
\caption{(a) Schematic of a single two-tone modulated cavity electromagnonic system in which a YIG sphere is placed inside a microwave cavity and biased by a two-tone modulated magnetic field. (b) Two YIG spheres are placed inside a microwave cavity and biased respectively by a constant and a two-tone modulated magnetic fields.}
\label{sys}
\end{figure}

As schematically depicted in Fig.\ref{sys} (a), we investigate a cavity electromagnonic system which consists of a YIG sphere inside the cavity. The spins in the YIG sphere interact with the magnetic field $\hat{\textbf{B}}$ from the electromagnetic field in the cavity and  externally applied uniform bias magnetic fields along the $z$ direction. We consider that the external magnetic field are two-tone modulated as
\begin{align}
B_{0t}=B_0+\sum_{j=1,2}B_j\cos(\nu _jt+\phi_j),
\end{align}
where $B_0$ is the strength of the constant component, $B_j$ the modulation depths, and $\nu_j$ the modulation frequencies, with the phases $\phi_j$. The interaction between the macrospin $\hat {\textbf{S}}$ in the sphere and the magnetic field (e.g. TE modes of the cavity) is described by
\begin{align}
\hat H_{\rm sb}&=-\gamma \hat {\textbf{B}}\hat{\cdot\textbf{S}},\nonumber\\
&=-\gamma B_{0t}\hat {S}_z+g_{\rm cs}(\hat c+\hat {c}^\dag)(\hat S_++\hat S_-),
\end{align}
with the gyromagnetic ratio $\gamma /2\pi =28$ GHz/T. Here we have let $\hat{\textbf{S}}=(\hat S_x, \hat S_y, \hat S_z )$ and the $x$-component $\hat S_x=(\hat S_++\hat S_-)/\sqrt{2}$ for the raising (lowering) operator $\hat S_+$ ($\hat S_-$) of spins. The parameter $g_{\rm cs}$ represents the coupling of the ensemble of spins to the cavity field denoted by the annihilation (creation) operator $\hat c$ ($\hat c^\dag$). Note that in the above derivation, the demagnetizing magnetic field and the anisotropic magnetic field
(caused by the magnetocrystalline anisotropy in the YIG sphere) are not taken into account since the nonlinear effects from them are weak such that they can be neglected. We have also assumed that the spins in the sphere experience the same magnetic-field strength of the electromagnetic field in the cavity since the diameter of the sphere is much smaller than the wavelength of the cavity microwave field. With the Holstein-Primakoff transformation \cite{en1}, the macrospin variables $\hat{S}_+=( \sqrt{2sN-\hat{m}^{\dagger}\hat{m}} ) \hat{m}$, $\hat{S}_-=\hat{m}^{\dagger}( \sqrt{2sN-\hat{m}^{\dagger}\hat{m}} )$, and $\hat S_z=sN-\hat{m}^{\dagger}\hat{m}$, in association with the bosonic annihilation (creation) operators $\hat{m} (\hat{m}^{\dagger})$, where the parameters $s$ and $N$ respectively represent single spin and the total number of spins in the YIG sphere. When the externally applied magnetic field is much stronger than that of the cavity field, the spins are almost aligned in the $z$ direction and we just investigate quantum fluctuations of the spins. Therefore, in such a
low-lying excitation situation, the excitation (magnon) of the collective spins is much smaller than the total spin number, i.e.,  $\langle \hat{m}^{\dagger}\hat{m} \rangle/2sN\ll 1$. Then the macrospin operators can be approximated as $\hat{S}^+\approx \sqrt{2sN}\hat{m}$, $\hat{S}^-\approx \sqrt{2sN}\hat{m}^{\dagger}$. The total Hamiltonian of the system is derived out as
\begin{align}
\hat{H_1}=\omega _c\hat{c}^{\dagger}\hat{c}+\omega _m\hat{m}^{\dagger}\hat{m}+g( \hat{c}+\hat{c}^{\dagger} ) ( \hat{m}+\hat{m}^{\dagger} )+\sum_{j=1}^2{\lambda _j\nu _j\cos ( \nu _jt+\phi_j )}\hat{m}^{\dagger}\hat{m},
\label{h2}
\end{align}
where $\omega_c$ is the resonant frequency of the cavity, the magnon frequency $\omega _m\equiv\gamma B_0$, $\lambda_j=\gamma B_j/\nu_j$,  and the cavity-magnon coupling strength $g=\frac{\gamma}{2} \sqrt{\frac{\hbar\omega_c\mu_0}{V_c} }\sqrt{2sN}$, with $V_c$ the mode volume of the microwave cavity resonance, $\mu _0$ the vacuum permeability ($\sqrt{\frac{\hbar\omega_c\mu_0}{V_c} }$: the vacuum amplitude of the magnetic field in the cavity)\cite{en2}.
Therefore, the external time-dependent driving magnetic field leads to the frequency modulation of the magnon mode in the YIG sphere. When just considering a time-independent bias magnetic field ($\lambda_j=0$) and for the single-photon coupling $g\ll \{\omega_c, \omega_m\}$, the above equation becomes approximately into 

\begin{align}
\hat{H_1}=\omega _c\hat{c}^{\dagger}\hat{c}+\omega _m\hat{m}^{\dagger}\hat{m}+g( \hat{c}\hat{m}^{\dagger}+\hat{c}^{\dagger}\hat m ),
\label{h3}
\end{align}
describing a beam-splitter-like interaction for nearly resonant frequencies of photons and magnons and only responsible for the exchange between photons and magnons\cite{Tian}.

Consider a rotating reference frame defined by the transformation operator $\hat{U}( t )=\hat{V}_1( t ) \hat{V}_2( t )$, where $\hat{V}_1( t )=\exp [ -i( \omega _s\hat{c}^{\dagger}\hat{c}+\omega _r\hat{m}^{\dagger}\hat{m} ) t ]$ and $\hat{V}_2( t ) =
\exp \{ -i[ \sum_{j=1}^2{\lambda _j\sin ( \nu _jt +\phi_j)} ] \hat{m}^{\dagger}\hat{m} \}$,
with  arbitrary frequencies $\omega _{s,r}$. In the rotating frame and with $e^{i\eta \sin x}=\sum_{z=-\infty}^{\infty}{J_z( \eta )}e^{izx}$, the Hamiltonian (\ref{h3}) becomes into

\begin{align}
&\hat{H}_{1I}=\delta _c\hat{c}^{\dagger}\hat{c}+\delta _m\hat{m}^{\dagger}\hat{m}+( g_{1t}\hat{m}+g_{2t}\hat{m}^{\dagger})\hat{c}+(g_{1t}\hat{m}^{\dagger}+g_{2t}\hat{m})\hat{c}^{\dagger},
\label{e1}
\end{align}
where the detuning $\delta _c=\omega _c-\omega _s$ and $\delta _m=\omega _m-\omega _r$, and the time-dependent coupling rates

\begin{subequations}
\begin{align}
&g_{1t}=g\sum_{z_1,z_2=-\infty}^{\infty}{J_{z_1}( \lambda _1 )}J_{z_2}( \lambda _2 ) e^{-i( \omega _s+\omega _r+z_1\nu _1+z_2\nu _2 ) t}e^{-i( z_1\phi _1+z_2\phi _2 )},\\
&g_{2t}=g\sum_{n_1,n_2=-\infty}^{\infty}{J_{n_1}( \lambda _1 )}J_{n_2}( \lambda _2 ) e^{-i( \omega _s-\omega _r-n_1\nu _1-n_2\nu _2 ) t}e^{i( n_1\phi _1+n_2\phi _2 )},
\end{align}
\end{subequations}
The Hamiltonian Eq.(\ref{e1}) is composed of rotating and counter-rotating terms with time-dependent coupling. We intend to only keep the nearly resonant terms  by choosing the modulation frequencies $\nu_j$.
When considering the conditions $\delta _{c,m}\lesssim g$ and choosing the modulation frequencies $\nu _1=\omega _r-\omega _s$ and $\nu _2=\omega _r+\omega _s$, such that
\begin{subequations}
\begin{align}
&\left| ( 1-z_1+z_2 ) \omega _s+( 1+z_1+z_2 ) \omega _r \right|\ll g,\\
&\left| ( 1+n_1-n_2 ) \omega _s-( 1+n_1+n_2 ) \omega _r \right|\ll g,
\end{align}
\end{subequations}
which can be satisfied by letting i.e., $z_1=0, z_2=-1$, $n_1=-1$  and $ n_2=0$, the Hamiltonian (\ref{e1}) is eventually changed into
\begin{align}
\hat{H}_{1I}=\delta _c\hat{c}^{\dagger}\hat{c}+\delta _m\hat{m}^{\dagger}\hat{m}
+( g_1\hat{m}e^{i\phi_2}+g_2\hat{m}^{\dagger}e^{-i\phi_1} )\hat{c}
+( g_1\hat{m}^{\dagger}e^{-i\phi_2}+g_2\hat{m}e^{i\phi_1} )\hat{c}^{\dagger},
\label{e2}
\end{align}
after discarding the nonresonant terms, with the time-independent coupling
\begin{subequations}
\begin{align}
g_{1t}\approx g_{1}= g J_0( \lambda _1 ) J_{-1}( \lambda _2 ),\\
g_{2t}\approx g_{2}= g J_{-1}( \lambda _1 ) J_0( \lambda _2 ).
\end{align}
\end{subequations}
The strengths $g_{1,2}$ can be adjusted by choosing the parameters $\lambda_{1,2}$. It is shown from (\ref{e2}) that the cavity field is coupled to the magnon mode via simultaneous parametric-downconversion and beam-splitter-like interactions, with unbalanced coupling strengths and tunable phase factors. It should be noted that when considering the situation $g_1\approx g_2=G$ (e.g., $G\approx-0.08g$ for $\lambda _1=\lambda _2=0.16$ ) and $\phi_1=\phi_2=\phi$, the above Hamiltonian
\begin{align}
&\hat{H}_{1I}=\delta _c\hat{c}^{\dagger}\hat{c}+\delta _m\hat{m}^{\dagger}\hat{m}
+G(\hat{m}e^{i\phi}+\hat{m}^{\dagger}e^{-i\phi} )(\hat{c}+\hat c^\dag),
\label{eee2}
\end{align}
reducing to the exact form of linearized cavity optomechanical coupling \cite{co}. Therefore, the modulated cavity electromagnonics can in principle exhibit the same quantum properties as demonstrated in cavity optomechanics. Further, if $\delta_{c,m}=0$ and $\phi=0$, the backaction-evading interaction between the magnons and photons
\begin{align}
&\hat{H}_{1I}=G(\hat{m}e^{i\phi}+\hat{m}^{\dagger}e^{-i\phi} )(\hat{c}+\hat c^\dag)
\label{eee3}
\end{align}
is achieved, which can be used for ultra-precision magnetic sensing \cite{ms1, ms2, ms3, ms4, ms5}, in analogy to weak force sensing via cavity optomechanics\cite{vco}. Note that in Ref.\cite{ms5}, such photon-magnon interaction is considered by fast modulating the single-photon coupling $g$ in Eq.(\ref{h2}), which is obviously difficult to realize experimentally.

\section{Steady magnonic squeezed states} 

At first, we consider the generation of steady-state magnonic squeezed state by exploiting the two-tone modulated cavity electromagnonics.  Using the approximate Hamiltonian (\ref{e2}) and taking into account of the cavity dissipation and magnon damping, the equations of motion for the cavity and magnon operator $\hat c$ and $\hat m$, given by

\begin{subequations}
\begin{align}
&\frac{d}{dt}\hat c=-\frac{\kappa_c}{2}\hat{c}-ig_1 \hat m^{\dagger}-ig_2\hat{m}+\sqrt{\kappa_c}\hat{c}^{\rm in}(t),\\
&\frac{d}{dt}\hat m=-\frac{\kappa_m}{2}\hat m-ig_1\hat c^\dag-ig_2\hat{c}+\sqrt{\kappa_m}\hat{m}^{\rm in}(t).
\end{align}
\label{ee3}
\end{subequations}
for $\phi_{1,2}=0$ and $\delta_{c,m}=0$, where $\kappa_c$ and $\kappa _m$ denote the rates of the cavity loss and magnonic damping, respectively. The operators $\hat{c}^{\rm in}(t)$ and $\hat{m}^{\rm in}(t)$ denote input noises coupled to the cavity field and magnon mode, satisfying zero mean and have nonzero correlations $\langle \hat{o}^{\rm in\dagger}( t ) \hat{o}^{\rm in}( t' ) \rangle=\bar{n}_o\delta ( t-t' )$ and $\langle \hat{o}^{\rm in}( t ) \hat{o}^{\rm in\dagger}( t' ) \rangle=( \bar{n}_o+1 )\delta ( t-t' )~(o=c,m \text{ and similarly hereinafter})$, where the equilibrium mean thermal excitations $\bar{n}_o=\big[ \exp ( \hbar \omega _o/k_BT ) -1 \big] ^{-1}$ at temperature $T$. By introducing the quadrature operators $u=( \hat{X}_c, \hat{P}_c, \hat{X}_m, \hat{P}_m ) ^T$, with $\hat{X}_o=( \hat{o}+\hat{o}^{\dagger} ) /\sqrt{2}$ and
$\hat{P}_o=-i( \hat{o}-\hat{o}^{\dagger} ) /\sqrt{2}$, the covariance matrix
$\Sigma_{cm}^{ij}=\langle  u_i u_j+u_j u_i \rangle /2-\langle u_i \rangle \langle u_j \rangle$ of the system is governed by 
\begin{align}
\frac{d}{dt}\Sigma_{cm}=A\Sigma_{cm}+\Sigma_{cm}A^T+D,
\label{cm}
\end{align}
with the matrices
\begin{align}
A=&-\left(
  \begin{array}{cccc}
	\frac{\kappa _c}{2}&		0&		0&		g_{1}-g_{2}\\
\\
	0&		\frac{\kappa _c}{2}&		g_{1}+g_{2}&		0\\
\\
	0&		g_{1}-g_{2}&		\frac{\kappa _m}{2}&		0\\
\\
	g_{1}+g_{2}&		0&		0&		\frac{\kappa _m}{2}\\
\end{array}
\right)
\label{m1}
\end{align}
and noise correlation matrix $D=\mathrm{diag}[ \kappa_c( \bar{n}_c+\frac{1}{2} ) , \kappa_c( \bar{n}_c+\frac{1}{2}), \kappa_m( \bar{n}_m+\frac{1}{2} ) , \kappa_m( \bar{n}_m+\frac{1}{2} ) ]$.

With the correlation matrix $\Sigma_{cm}$, we can study the properties of the squeezing of the magnon mode and cavity field. The optimal squeezing of the magnon and cavity modes can be quantified by
\begin{align}
V_o=\mathrm{Min}\big\{ \mathrm{Eigenvalue}[ \mathrm{\Sigma}_{m,c} ]\big\},
\end{align}
where $\Sigma_{m,c}$ is the reduced correlation matrices of the magnon mode and cavity field. $V_o<1/2$ indicates the squeezing and the smaller values mean stronger squeezing.

\begin{figure}[t]
\centerline{\scalebox{0.65}{\includegraphics{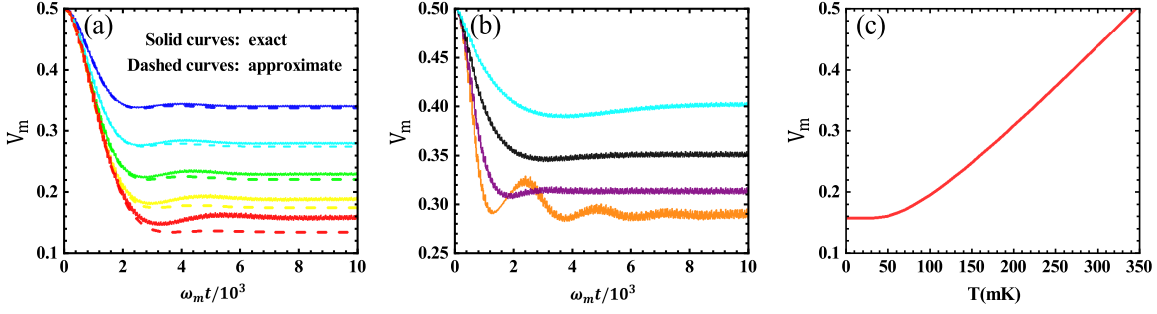}}}
 \caption{(a) Time evolution of the magnon squeezing $V_{m}$ for the exact (solid curves) and approximate (dashed curves) results, with
$\lambda _1=0.2$,  $\kappa _c =2\times10^{-3}\omega_m$, $\kappa _m =2\times10^{-5}\omega_m$, $\omega _c =0.45\omega_m$, $g =10^{-2}\omega_m$, $T=0$, and $\lambda _2=0.04$ (blue), $0.06$ (cyan), $0.08$ (green), $0.1$ (yellow), and $0.12$ (red)  [corresponding to $g_1/g_2=0.2, 0.3, 0.4, 0.5$ and $0.6$, respectively].
 (b) The magnon squeezing $V_{m}$ for the exact results, with $\lambda _1=0.2$, $\lambda _2=0.06$, $\kappa _m =2\times10^{-4}\omega_m$, $g =1.5\times10^{-2}\omega_m$, $\omega _c =0.45\omega_m$, $T=0$,  $\kappa _c =10^{-3}\omega_m$ (orange), $4\times10^{-3}\omega_m$ (violet),  $10^{-2}\omega_m$ (black), $2\times10^{-2}\omega_m$ (cyan).
 (c) The effect of temperature  on the long-time magnon squeezing $V_{m}$ for $\lambda _1=0.2$ and $\lambda _2=0.12$, and the other parameters are the same as in (a).}
 \label{f1}
\end{figure}

In Fig.\ref{f1}, we present the exact results [solid curves, obtained with the full Hamiltonian (\ref{h2})] and the approximate results [dashed curves, obtained with the approximate Hamiltonian (\ref{e2})] for the variance $V_m$ by numerically solving Eq.(\ref{cm}) and considering the modulation frequency $\nu_1=\omega_m-\omega_c$ and $\nu_2=\omega_m+\omega_c$. It is clearly shown that the magnon squeezing can indeed be generated via the two-tone modulation. For instance, the steady-state magnon squeezing $V_{m}\approx0.14$ for the ratio $\lambda_1/\lambda_2\approx1.67$, as shown from Fig.\ref{f1}~(a).
With the increasing of $\lambda_1/\lambda_2$ increases, the squeezing grows up obviously. Fig.\ref{f1}~(b) shows that for the fixed magnon damping rate $\kappa_m$, the steady magnon squeezing is enhanced when the cavity loss rate $\kappa_c$ arises, and compared with Fig.\ref{f1}~(a), its degree decreases since the magnon damping rate increases. We can see that the approximate results is closes to the exact results. In fact, by introducing  new operator
 $\hat {\tilde{m}}=\sinh r\hat{m}^{\dagger}+\cosh r\hat{m}$
 with $[\hat {\tilde{m}}, \hat {\tilde{m}}^\dag]=1$, which corresponds to the unitary
 transformation
$\hat{\tilde{m}}=\hat{S}( r ) \hat{m}\hat{S}( -r )$,
with the squeezing operator $\hat{S}( r )=\exp [ r( \hat{m}^2-\hat{m}^{\dagger 2} ) /2 ] $
and squeezing parameter  $r=\tanh ^{-1}( g_1/g_2 )$.
In the transformed picture, the Hamiltonian ($\ref{e2}$) becomes into
\begin{align}
\tilde{H_I}=\Omega_1 ( \hat{\tilde{m}}^{\dagger}\hat{c}+\hat{\tilde{m}}\hat{c}^{\dagger} ),
\end{align}
where $\Omega_1=\sqrt{g_{2}^{2}-g_{1}^{2}}$.
\begin{figure*}[t]
\centerline{\scalebox{0.65}{\includegraphics{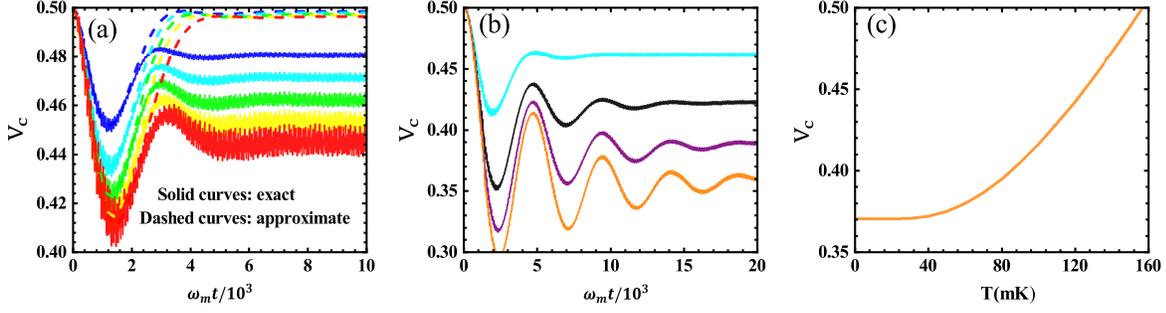}}}
 \caption{(a) Time evolution of the microwave-field squeezing $V_{c}$ for the exact and approximate results, with the parameters
$\lambda _1=0.2$, $\kappa _c =2\times10^{-3}\omega_m$, $\kappa _m =2\times10^{-5}\omega_m$, $\omega _c =0.45\omega_m$, $g =10^{-2}\omega_m$, $T=0$, and $\lambda _2=0.04$ (blue), $0.06$ (cyan), $0.08$ (green), $0.1$ (yellow), and $0.12$ (red).
 (b) The cavity field squeezing $V_{c}$  for the exact results, with $\lambda _1=0.2$, $\lambda _2=0.06$, $\kappa _m =2\times10^{-4}\omega_m$, $g=0.7\times10^{-2}\omega_m $, $\omega _c=0.45\omega_m $, $T=0$, and $\kappa _c =2\times10^{-3}\omega_m$ (cyan), $4\times10^{-4}\omega_m$ (black), $2\times10^{-4}\omega_m$ (violet), and $\kappa _c=5\times10^{-5}\omega_m$ (orange).
 (c) The effect of temperature on the microwave-field squeezing $V_{c}$, for $\lambda _1=0.2$, $\lambda _2=0.06$, $\kappa _c =5\times10^{-5}\omega_m$, $\kappa _m =2\times10^{-4}\omega_m$ and the other parameters are the same as in (b).}
 \label{f2}
\end{figure*}
For $\kappa_c\gg\kappa_m$ to neglect the magnon damping, the modes $\hat c$ and $\hat {\tilde{m}}$ will be in vacuum in the long-time regime,
i.e., the mode  $\hat {\tilde{m}}$ is cooled to the ground state. In the original picture the magnon mode $\hat m$ is therefore prepared in squeezed vacuum, i.e., $|\psi_m\rangle_{\rm ss}=\hat S(-r)|0_m\rangle$.
Consequently, the increasing of $\lambda_1/\lambda_2$ enhances the steady-state squeezing, 
as the squeezing parameter $r$ (or $g_1/g_2$) is increased. In addition, as the cavity loss rate $\kappa_c$
arises, the condition $\kappa_c\gg\kappa_m$ is better satisfied, and thus the squeezing accordingly increases,
shown in Fig.\ref{f1}~(b). Obviously, with finite magnon damping rate, in the steady-state regime
the cavity is not in vacuum, but also in squeezed states, as shown from Fig.\ref{f2}.
Since we have assumed $\kappa_m\ll\kappa_c$, the squeezing degree of the cavity field is much smaller than
that of the magnon mode. If we decrease the cavity dissipation rate $\kappa_c$, stronger microwave squeezing can also be obtained, as demonstrated in Fig.\ref{f2}~(b). The effects of thermal
environment on the squeezing are plotted In Fig.\ref{f1}~(c) and Fig.\ref{f2}~(c). We see that increase in temperature suppresses
the generation of the steady squeezed states of the magnon and cavity mode.
With the increase of temperature, the squeezings decrease and eventually disappear. The magnonic squeezing is
robust against thermal fluctuations and can exist up to the temperature above 300 mK.

\section{Steady magnonic entangled states}

We next study the direct generation of the steady-state entanglement between magnon modes
in two YIG spheres by a two-tone modulated cavity electromagnonics. As depicted in Fig.\ref{sys} (b),
we consider a tripartite cavity electromagnonic system which consists of two YIG spheres
inside a microwave cavity. We further consider that one YIG sphere is driven
by a time-modulated external magnetic field, as in Fig.1~(a),
and the other one is just driven by a time-independent bias magnetic field.
The magnon modes in the two YIG spheres are simultaneously coupled to the cavity field.
The total Hamiltonian of the tripartite system can be written as
\begin{align}
\hat{H_2}= &\omega _c\hat{c}^{\dagger}\hat{c}+\sum_{j=1}^2\omega _{m_j}\hat{m}_j^{\dagger}\hat{m}_j
+\sum_{j=1}^2{\lambda _j\nu _j\cos ( \nu _jt )}\hat{m}_1^{\dagger}\hat{m}_1\nonumber\\
&+g( \hat{c}+\hat{c}^{\dagger} ) ( \hat{m}_1+\hat{m}_1^{\dagger} )+g_3( \hat{c}+\hat{c}^{\dagger} ) ( \hat{m}_2+\hat{m}_2^{\dagger} ),
\label{e4}
\end{align}
where $\hat{m}_j~(\hat{m}_j^{\dagger})$ is the annihilation (creation) operators of the $j$th magnon mode of frequency $\omega_{m_j}$.
The strength $g$ and $g_{3}$ represent the interaction between the cavity field and the magnon modes via the magnetic dipole interaction.
With the same procedure as in Sec.II,
the Hamiltonian (\ref{e4}) can be expressed as
\begin{align}
&\hat{H}_{2I}= ( g_{1t}\hat{m}_1+g_{2t}\hat{m}_1^{\dagger} ) \hat{c}+( g_{1t}\hat{m}_1^{\dagger}+g_{2t}\hat{m}_1) \hat{c}^{\dagger}\nonumber\\
&~~~~~~+g_3( \hat{m}_2\hat{c}^{\dagger}+\hat{m}_2^{\dagger}\hat{c}+\hat{m}_2\hat{c}e^{-2i\omega_{m_2}t}+\hat{m}_2^{\dagger}\hat{c}^{\dagger}e^{2i\omega_{m_2}t} ).
\label{e5}
\end{align}
Similarly, choosing the time-dependent couplings
$g_{1t}\approx g_1=g J_0( \lambda _1 ) J_{-1}( \lambda _2 )$ and $g_2\approx g_{2t}=g J_{-1}( \lambda _1 ) J_0( \lambda _2 )$
and further adjusting the parameters $\lambda_j$
such that the coupling $g_1 \ne 0$ and $g_2 \approx0$ (e.g., $\lambda _1=3.817$ and $\lambda _2=0.3$), the Hamiltonian (\ref{e5}) reduces to
\begin{align}
\hat{H}_{2I}\approx g_1( \hat{m}_1\hat{c}+\hat{m}_1^{\dagger}\hat{c}^{\dagger} )+g_3( \hat{m}_2\hat{c}^{\dagger}+\hat{m}_2^{\dagger}\hat{c} ).
\label{e6}
\end{align}
for $\omega_c=\omega_{m2}$ and $g_3\ll\omega_c$. The first part describes an effective parameter downconversion of the microwave field and the first mangon mode $\hat{m}_1$, which leads to entanglement between them, while the second part describes a beam-splitter-like interaction between the second magnon mode $\hat{m}_2$ and the cavity field, which may transform the entanglement built up between $\hat {m}_1$ and $\hat c$ into the entanglement between the two magnon modes. When taking into account of the cavity and magnon losses, the equations of motion of the sysytem's operators are derived as
\begin{subequations}
\begin{align}
&\frac{d}{dt}\hat c=-\frac{\kappa_c}{2}\hat{c}-ig_{1} \hat{m}_1^{\dagger}-ig_{2} \hat{m}_1-ig_3\hat{m}_2+\sqrt{\kappa_c}\hat{c}^{\rm in}(t),\\
&\frac{d}{dt}\hat m_1=-\frac{\kappa_{m_1}}{2}\hat{m}_1-ig_{1} \hat{c}^{\dagger}-ig_{2} \hat{c}+\sqrt{\kappa_{m_1}}\hat{m}_1^{\rm in}(t),\\
&\frac{d}{dt}\hat m_2=-\frac{\kappa_{m_2}}{2}\hat{m}_2-ig_{2}\hat{c}-ig_3 \hat{c}+\sqrt{\kappa_{m_2}}\hat{m}_2^{\rm in}(t).
\end{align}
\end{subequations}
where $\kappa_{m_j}$ denotes the damping rate of the $j$th magnon mode and the input noise operator $\hat{m}_j^{\rm in}(t)$ satisfies the nonzero correlations
$\langle\hat{m}_j^{\rm in}( t ) \hat{m}_j^{\rm in\dagger}( t' ) \rangle=( 1+\bar{n}_j ) \delta ( t-t' )$ and $\langle\hat{m}_j^{\rm in \dagger}( t ) \hat{m}_j^{\rm in}( t' ) \rangle=\bar{n}_j  \delta ( t-t' )$, with $\bar{n}_j=\left[ \exp ( \hbar \omega _{m_j}/k_BT ) -1 \right] ^{-1}$ and temperature $T$. By defining the quadrature operators $\tilde u =( \hat{X_c}, \hat{P_c}, \hat{X_1}, \hat{P_1}, \hat{X_2}, \hat{P_2}) ^T$, with $\hat{X}_j=( \hat{m}_j+\hat{m}_j^{\dagger} ) /\sqrt{2}$,
$\hat{P}_j=-i( \hat{m}_j-\hat{m}_j^{\dagger} ) /\sqrt{2}$, the covariance matrix
$\Sigma_{cm_{12}}^{ij}=\langle  \tilde u_i\tilde u_j+\tilde u_j \tilde u_i \rangle /2-\langle \tilde u_i \rangle \langle \tilde u_j \rangle$ of the tripartite system satisfies the equation of the same form as Eq.(\ref{cm}), just with the replacement by

\begin{align}
A=&\left(
  \begin{array}{cccccc}
      -\frac{\kappa _c}{2} & 0 & 0 & -g_{1}+g_{2} & 0 & g_3\\
     \\
     0 & -\frac{\kappa _c}{2} & -g_{1}-g_{2} & 0 & -g_3 & 0 \\
     \\
    0 & -g_{1}+g_{2} &  -\frac{\kappa _{m_1}}{2} &  0 & 0 & 0\\
     \\
    -g_{1}-g_{2} & 0 & 0 & -\frac{\kappa _{m_1}}{2} & 0 & 0\\
     \\
    0 & g_3 & 0 & 0 &-\frac{\kappa _{m_2}}{2}  & 0\\
     \\
     -g_3& 0 & 0 & 0 & 0 & -\frac{\kappa_{m_2}}{2}\\
 \end{array}
\right),
\end{align}
and $D=\mathrm{diag}[ \kappa _c ,\kappa _c , \kappa_{m_1}( \bar{n}_1+\frac{1}{2} ) , \kappa_{m_1}( \bar{n}_1+\frac{1}{2} ) , \kappa_{m_2}( \bar{n}_2+\frac{1}{2} ) , \kappa_{m_2}( \bar{n}_2+\frac{1}{2} ]$.

\begin{figure}[t]
\centerline{\scalebox{0.65}{\includegraphics{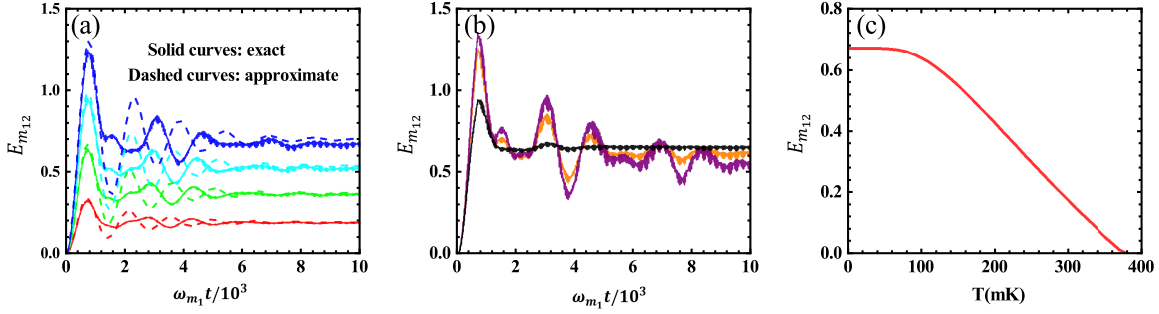}}}
 \caption{(a) Time evolution of the magnonic entanglement $E_{m_{12}}$ for the exact and  approximate results, with $\lambda _1=3.8317$ and $\lambda _2=0.075$ (red), $0.15$ (green), $0.225$ (cyan), and $0.3$ (blue) [corresponding to the coupling ratio $g_1/g_3=0.1, 0.2, 0.3, 0.4$, respectively].
The other parameters $\kappa _{m_1}=\kappa_{m_2}=2\times10^{-5}\omega_{m_1}$, $\kappa _c =2\times10^{-3}\omega_{m_1}$, $\omega _{m_2} =\omega _c =0.85\omega_{m_1}$, $g =3\times10^{-2}\omega_{m_1}$, $g_3 =4.5\times10^{-3}\omega$, and $T=0$.
 (b) The entanglement $E_{m_{12}}$, for $\lambda _1=3.8317$, $\lambda _2=0.3$, $\kappa _{m_1}=\kappa _{m_2} =2\times10^{-4}\omega_{m_1} $, $\kappa _c=4\times10^{-4}\omega_{m_1}$ (violet), $10^{-3}\omega_{m_1}$ (orange), and
$4\times10^{-3}\omega_{m_1} $ (black). (c) The effect of temperature  on the entanglement $E_{m_{12}}$ for $\lambda _1=3.8317$ and $\lambda _2=0.3$. The other parameters in (b) and (c) are the same as in (a).}
 \label{f3}
\end{figure}

\begin{figure}[t]
\centerline{\scalebox{0.8}{\includegraphics{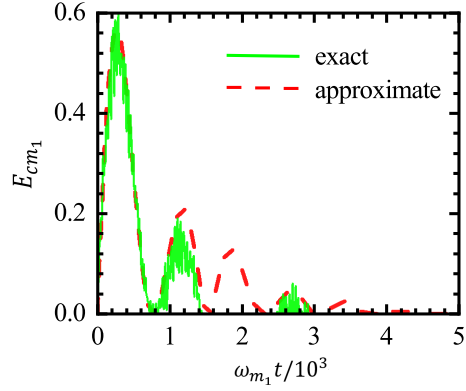}}}
 \caption{Time evolution of the entanglement $E_{cm_1}$ between the cavity field $\hat{c}$ and the magnon mode $\hat m_1$ for the exact and approximate results, for the parameters $\kappa _{m_1} =\kappa _{m_2} =2\times10^{-5}\omega_{m_1}$, $\kappa _c =2\times10^{-3}\omega_{m_1}$, $\omega _{m_2}=\omega _c=0.85/\omega_{m_1}$, $g=3\times10^{-2}\omega_{m_1}$, $g_3 =4.5\times10^{-3}\omega_{m_1}$, $\lambda _1=3.8317$, $\lambda _2=0.3$ and $T=0$.}
 \label{f4}
\end{figure}
The bipartite entanglement in the tripartite system characterized can be quantified by the logarithmic negativity \cite{en3}
\begin{equation}
E_{m_{12}(cm_1)}=\max\big[0,-\log(2\tilde{\nu}^-_{m_{12}(cm_1)})\big],
\label{e19}
\end{equation}
where $\tilde{\nu}^-_{m_{12}(cm_1)}$ is the smallest symplectic eigenvalues of the partially-transposed reduced correlation matrices $\Sigma_{m_{12}}$ of the two magnon modes and $\Sigma_{cm_1}$ of the magnon mode $\hat m_1$ and the cavity field.

The exact results, obtained with the full Hamiltonian (\ref{e4}), are presented in Fig.\ref{f3} (solid curves). We see from it that the magnonic entanglement can be achieved in both transient and steady-state regimes. The long-time entanglement $E_{m_{12}}\approx 0.7$ for the parameters $\lambda_2=0.3$. With the fixed value of $\lambda_1=3.8317$ at which the coupling $g_2\approx 0$, the increasing of $\lambda_2$ enhances the steady-state entanglement since the coupling $g_1$ also grows up accordingly, shown in Fig.\ref{f3}~(a). It is shown from Fig.\ref{f3}~(b) that the entanglement in the long-time regime increases as the cavity loss rate $\kappa_c$ arises. In addition, the approximate results, obtained with the Hamiltonian (\ref{e6}), are also close to the exact ones in the long-time regime.
In fact, if we introduce a new bosonic operators $\hat {\tilde{m}}_1=( \sinh r_2 \hat{m}_1^\dag+\cosh r_2\hat{m}_2 )$ and $\hat {\tilde{m}}_2=( \sinh r_2 \hat{m}_2^\dag+\cosh r_2\hat{m}_1 )$, corresponding to the two-mode squeezing transformation $\hat {\tilde{m}}_{1,2}=\hat S_{12}(r_2)\hat m_{1,2}\hat S_{12}(-r_2)$, with the squeezing operator $\hat S_{12}(r_2)=\exp[r_2(\hat m_1\hat m_2-\hat m_1^\dag\hat m_2^\dag)]$ and squeezing parameter with $r_2=\tanh^{-1}(g_1/g_3)$,
then the Hamiltonian equation Eq.(\ref{e2})
\begin{align}
\hat {\tilde{H}}_{2I}=\Omega_2 ( \hat {\tilde{m}}_1^{\dagger}\hat{c}+ \hat {\tilde{m}}_1\hat{c}^{\dagger}),
\label{e3}
\end{align}
with the coupling $\Omega_2 =\sqrt{g_{3}^{2}-g_{1}^{2}}$. It is shown that in the transformed picture, only the mode $\hat {\tilde{m}}_1 $ is coupled to the cavity mode, while the other mode $\hat {\tilde{m}}_2$ is decoupled to the system. When neglecting the magnon damping for $\kappa_c\gg\kappa_m\equiv \kappa_{m_j}$, the cavity dissipation drives the cavity field and the mode $\hat {\tilde{m}}_1$ in vacuum in the long-time limit, leaving the mode $\hat {\tilde{m}}_2$ always in a thermal state $\hat \rho_{\tilde m_2}$ with average thermal number $\bar {n}_{\tilde m_2}=\sinh^2 r_2$. Therefore, in the original picture, the two magnon modes are eventually driven into a two-mode squeezed thermal state, i.e.,
\begin{align}
\rho_{m_{12}}^{\rm ss}=\hat S_{12}(-r_2)|0\rangle\langle 0| \otimes \hat \rho_{\tilde m_2}\hat S_{12}^\dag(-r_2).
\end{align}
Hence, the steady-state magnonic entanglement increases with the increasing of $g_1$ and $\kappa_c$. The entanglement can still exist even when the temperature $T=300$ mK. In Fig.\ref{f4}, we see that the bipartite entanglement $E_{cm_1}$ between the cavity field $\hat c$ and the magnon $\hat m_1$ appears just in the transient regime since in the long-time regime the cavity field is almost dissipated into vacuum.

\section{Conclusion}
In conclusion, we propose a scheme to generate the steady squeezed and entangled magnonic states via two-tone modulated cavity electromagnonics. Through the modulation, an effective Hamiltonian exactly same as that of generic linearized cavity optomechanics can be formed, which allows us to engineer nonclassical magnonic states directly via cavity electromagnonics, in analogy with the cavity optomechanics. The modulation can also be  used to realize back-action-evading measurement of magnonic amplitude  for ultrasensitve weal-magnetic sensing. Hence, the present scheme provides promising opportunities for the preparation of macroscopic quantum states of collective spin excitations in solids and magnon-based quantum information processing and metrology.

\ack

This work is supported by the National Natural Science Foundation of China (No. 12174140).

\section*{Data availability statement}
The data that support the findings of this study are available upon reasonable request from the authors.

\end{document}